\documentclass[aps,prb,twocolumn,superscriptaddress,floatfix,amsmath,amssymb]{revtex4-1}

\usepackage{graphicx}
\usepackage{bm}
\usepackage[colorlinks=true,urlcolor=blue,citecolor=blue,linkcolor=blue]{hyperref}

\begin{document}

\title{Interpretable Machine Learning Study of Many-Body Localization Transition in Disordered Quantum Ising Spin Chains}

\author{Wei Zhang}
\affiliation{Department of Physics, Boston College, Chestnut Hill, MA 02467,
USA}
\author{Lei Wang}
\affiliation{Beijing National Lab for Condensed Matter Physics and Institute
of Physics, Chinese Academy of Sciences, Beijing 100190, China}
\author{Ziqiang Wang}
\affiliation{Department of Physics, Boston College, Chestnut Hill, MA 02467,
USA}

\date{\today}

\begin{abstract}
We apply support vector machine (SVM) to study the phase
transition between many-body localized and thermal phases in a
disordered quantum Ising chain
in a transverse external field. The many-body eigenstate energy $E$ is bounded by a bandwidth $W=E_{max}-E_{min}$. The transition takes place on a phase diagram spanned by the energy density $\epsilon=2(E-E_{min})/W$ and the disorder strength $\delta J$ of the spin interaction uniformly distributed within $[-\delta J, \delta J]$, formally parallel to the mobility edge in Anderson localization. In our study we use the labeled probability density of eigenstate wavefunctions belonging to the deeply localized and thermal regimes at two different energy densities ($\epsilon$'s) as the training set, {\em i.e.}, providing labeled data at four corners of the phase diagram. Then we
employ the trained SVM to predict the whole phase diagram.  The obtained phase boundary qualitatively agrees with previous work using entanglement entropy to
characterize these two phases. We further analyze the decision function of
the SVM to interpret its physical meaning and find that it is analogous to the inverse participation ratio in configuration space. Our findings demonstrate the ability of the SVM to capture potential quantities that may characterize the many-body localization phase transition.

\end{abstract}

\maketitle


\section{\label{sec:level1}Introduction}

Many-body localization (MBL) refers to a class of correlated systems that fail to
thermalize in the sense that they violate the eigenstate thermalization
hypothesis (ETH) \cite{Deutch_1991,Srednicki_1994,Rigol_2008,Nandkishore_2015}. As a consequence, certain memories of the local initial conditions can be forever remembered in
conserved local observables. They thus have the potential to robustly store quantum information \cite{Huse_2013}. Compared to the conventional thermal phase, the MBL phase has many novel characteristic properties. The hallmark of the MBL phase is that the eigenstate entanglement entropy follows the
area-law instead of the volume-law in the thermal
phase \cite{Kjall_2014,Luitz_2015,Bauer_2013,Znidaric_2008,
Bardarson_2012,Vosk_2013,Serbyn_2013_1,Serbyn_2013_2,Grover,Eisert_2010}.
The MBL phase has zero DC conductivity \cite{Bakso_2006} and discrete local spectrum \cite{Nandkishore_2014}.
The statistics of the energy level spacing in the MBL phase is described by the Poisson distribution, in contrast to the Wigner-Gaussian distribution typical in the thermal phases \cite{Huse_2013,Luitz_2015,Nandkishore_2015,Geraedts_2016,
Serbyn_2013_2,Swingle,Ros_2015}.

The properties of the entanglement entropy and the level spacing have been commonly used to study MBL-thermal phase transition \cite{Kjall_2014,Luitz_2015,Shem_2015,Modak_2015,Baygan_2015,Luitz_2016,Nag_2017}. However, the intrinsic many-body problem makes the study of the
critical phenomena very challenging due to the sample size limitations and the nonperturbative nature of strong disorder. Despite the formal analogy to the mobility edge problem in the single particle Anderson localization \cite{Anderson_1958}, such basic questions of whether the MBL-ETH transition can be viewed as a localization transition in the many-body Hilbert space remains controversial. It is known that Anderson localization is stable against weak electron-electron interactions, which suggests that the MBL phase would emerge when disorder is strong enough \cite{Bakso_2006}. One of the most profound and powerful physical quantities widely used to identify the Anderson localization transition is the inverse participation ratio (IPR) \cite{Brndiar_2006} that measures the (inverse) of the spatial coverage of the single-particle eigenstates. One therefore asks if the MBL arises through the localization of the many-body states in the configurational Hilbert space, and if the scaling behavior of properly generalized IPR can be used to determine the MBL phase transition.
Several theoretical studies have shown that the behavior of the IPR (or its inverse) and the entanglement entropy
share similarities \cite{Torres_2016,Bera_2015,Beugeling_2015} and are directly related in the single particle picture \cite{Chen_2012}, whereas others offer opposite arguments \cite{Luitz_2015,Biroli_2010}. Recent experimental measurements also explored and demonstrated the connections between Hilbert space localization and energy level statistics\cite{Roushan_2017}.

In this work, we apply machine learning to the classification of two
different phases, the ETH and the MBL. We will also explore and extract useful information concerning the above questions from a machine learning perspective. Specifically, we build and operate the support vector machine (SVM), designed for the random transverse-field Ising chain.
First, we demonstrate that the trained
SVM with appropriate kernel choice is able to distinguish the two phases and determine the phase boundary. For our model, we only require training data from two different energy densities to make the trained SVM work for the whole energy spectrum. This fact ensures that during the training process, the models are built on properties of the MBL phase itself which should not depend on energy. Compared to training and testing at a fixed energy density and repeat the process multiple times in the full energy space to determine the transition line, training only once is much more computation cost-saving, especially considering that it is often expensive to generate class labels. Finally, we try to study and understand how the SVM makes the decision. We find strong evidence that the SVM has the ability to automatically choose a decision function which is very closely related to the many-body IPR defined in the configuration space.

\section{\label{sec:level2}Model and Method}

\subsection{Transverse-field disordered quantum Ising chain}
\begin{figure}[h!]
 \includegraphics[trim = 0mm 0mm 0mm 0mm,width=1\columnwidth,clip=true]{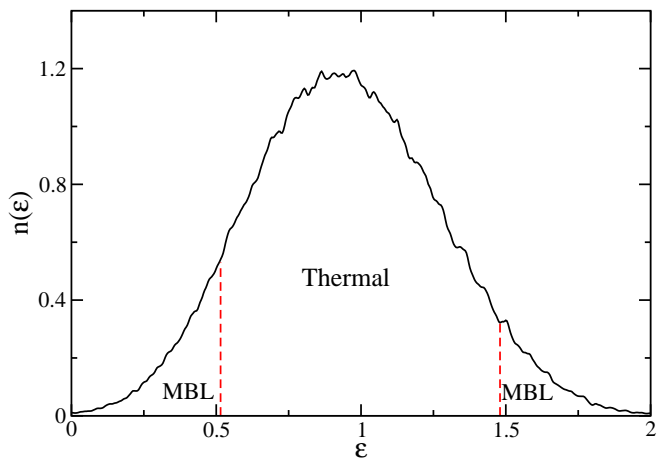}
\caption{Density of state of Hamiltonian in Eq.~(\ref{equation:Hamiltonian}) at $\delta J=1.8$ for a specific disorder configuration. $\epsilon$ is the energy density. The mobility edges separating thermal and MBL phases are determined according to supplementary material of \cite{Kjall_2014}.}
\label{fig:DoS}
\end{figure}

The quantum transverse-field Ising chain is known to develop the MBL phase when the disorder strength is strong. The Hamiltonian of the system is given by~\cite{Kjall_2014}
\begin{equation}
\hat{H}=-\sum_{i=1}^{L-1}J_i\sigma_i^z\sigma_{i+1}^z+
J_2\sum_{i=1}^{L-2}\sigma_i^z\sigma_{i+2}^z+h\sum_{i=1}^L\sigma_i^x
\label{equation:Hamiltonian}
\end{equation}
where $\sigma^{x}$ and $\sigma^{z}$ are Pauli matrices and $L$ is the number of sites in the chain. In Eq.~(\ref{equation:Hamiltonian}), the second nearest neighbor coupling $J_2$ and the transverse eternal field $h$ will be assigned uniform and nonrandom values, whereas the nearest neighbor coupling is site-dependent, $J_i=J+\delta J_i$, where $J$ is a constant and $\delta J_i$ is randomly taken from a uniform distribution $[-\delta J, \delta J]$. Thus $\delta J$ measures the
disorder strength. For a certain disorder realization, the energy $E$ of the many-body eigenstates of $H$ is bounded within a bandwidth $W= E_{max}-E_{min}$. Consider a disordered ensemble of $H$, the appropriate dimensionless energy is defined by the energy density $\epsilon=2(E-E_{min})/W$ relative to the total bandwidth, within a small window around $\epsilon$. The density of states of this model at $\delta J=1.8$ when $L=14$ for a specific disorder configuration is shown in Fig.~\ref{fig:DoS}. For a given set of $J$, $J_2$, and $h$, the transition between the thermal (ETH) and MBL phases corresponds to a boundary in the phase diagram spanned by $\delta J$ and $\epsilon$. Here we set $J_2=0.5h=0.3 J$.

We express the many-body quantum states and the Hamiltonian matrix in the spin configuration basis, which is constructed by direct product states of the {\em local} Hilbert space $\{\sigma_i^z\}$. In addition to being natural, this basis is non-entangled and suitable for introducing the many-body IPR to describe the localization in the spin configuration basis. We work in this basis throughout the rest of the paper.

\subsection{Data for machine learning}

Instead of dividing the system into two subsystems $A$ and $B$ to
calculate the reduced density matrix of an eigenstate
$\rho_A={\rm Tr}_B|\Psi\rangle\langle\Psi|$ and using
the entanglement spectrum as the training data set \cite{Schindler_2017,Nieuwenburg_2017}, we directly feed the
probability density of the eigenstate $|\Psi\rangle$ computed in the spin basis to the machines as the training data set.
The reason for doing so is that, although by preprocessing the training data can reduce the dimension and filter out redundant information, useful information contained in the wavefunction of the entire system can also be lost. Since the entanglement entropy is not the only quantity that can characterize the MBL phase, we thus classify the probability density of the wave function instead of the entanglement spectra. This method not only allows the exploration of other characteristic physical quantities of MBL in the entire system, but also stages a test on the power of machine learning: if only the minimally processed knowledge is provided in the training data, will machine learning be able to find out the relevant physical property to be used for classification by itself?

Our results show that the answer is affirmative. In addition, the algorithm turns out to be remarkably efficient for our model: only input wave functions at two different energy densities are used as the training set and the trained model is able to determine the transition region at all energy densities and the mobility edge for any disorder strength. In other words, by training with wave functions generated at four corner points on the $(\delta J, \epsilon)$-plane, the models are able to produce the complete phase boundary in the 2-parameter phase diagram. It is also remarkable that the SVM is capable of capturing certain generic properties for all energy densities in making the decision, rather than being trapped by energy-specific properties. This part is presented in detail together with the classification results and the decision function detection in Section
\uppercase\expandafter{\romannumeral3}.

\subsection{Support vector machine}

There are many machine learning models that are widely used for data classification. Some of them have been applied to study phase transitions in many-body systems, such as artificial neural networks~\cite{Carrasquilla_2017,Nieuwenburg_2017,Schindler_2017,Zhang_2017,Hsu_2018}, clustering via principal component analysis~\cite{Wang_2016}, and kernel method for support vector machine (SVM)~\cite{Ponte_2017,Greitemann_2018}. Here we focus on the last one due to its better interpretability.

\begin{figure}[h!]
 \includegraphics[trim = 0mm 0mm 0mm 0mm,width=1\columnwidth,clip=true]{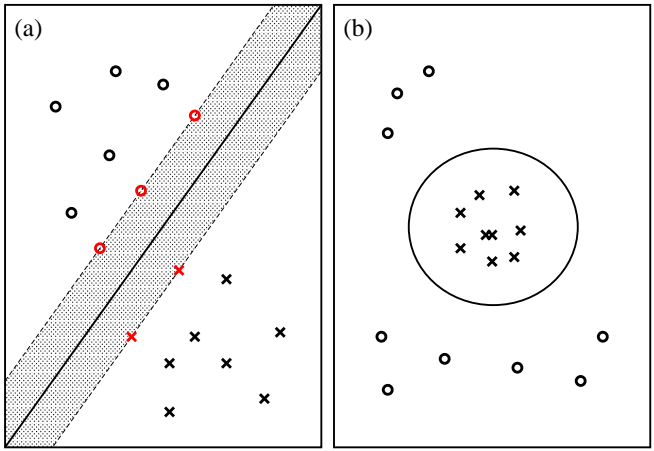}
\caption{(a) A separating plane (solid line) separates two different phases (labeled as circles and crosses respectively) with the largest margin (shaded area) in the 2-dimensional feature space. The red circles and crosses mark the support vectors that are closest to the separating plane.  (b) The large circle in the original 2-dimensional feature space is a separating hyperplane in higher dimensional space after the transformation. Such a transformation makes data points that are not linearly separable in its original space linearly separable in the transformed higher dimensional space.}
\label{fig:schematic}
\end{figure}

SVM is one of the most successful model for binary classification, which aims to linearly separate data belonging to two classes $\{+1,-1\}$, making the distance between the separating hyperplane and its nearest data points in both classes as large as possible. In other words, for any hyperplane separating the two classes of data, there exists a region where we can pin the separating hyperplane without changing the accuracy of classification. This region is called the margin and we want to find the hyperplane corresponding to the maximum margin. Fig.~\ref{fig:schematic}(a) is a schematic plot of how a separation plane separates different phases with largest margin in a two-dimensional feature space.

The hyperplane satisfying this requirement can be described by the linear equation: $\vec{w}\cdot \vec{x}+b=0$, where $\vec{w}$ is the vector perpendicular to the hyperplane and $\vec{x}$ denotes any point on the hyperplane. Since only the direction of $\vec{w}$ matters,
we can rescale the modulus of $\vec{w}$ and make the distance between the separating hyperplane and its closest data points equal to one. Denoting those data points closest to the separating hyperplane as $\vec{x}^{SV}$ (where the superscript SV stands for support vectors), we have,
after rescaling, $|\vec{w}\cdot \vec{x}^{SV}+b|=1
\label{equation:normalize}$.
As a result, the distance from $\vec{x}^{SV}$ to the hyperplane,  $|(\vec{x}^{SV}-\vec{x})\cdot\frac{\vec{w}}
{|\vec{w}|}|=\frac{1}{|\vec{w}|}$ is what we want to maximize. Equivalently, we can minimize $\frac{1}{2}\vec{w}\cdot\vec{w}$ subject to the condition $y_n(\vec{w}\cdot\vec{x}_n+b)\geq 1$, where $\vec{x}_n$ is any of the training data samples in the two classes $y_n=\pm 1$, because the distance from any of them to the separating hyperplane is at least $1$.

Next, consider the case where the data points are not completely linearly separable, i.e. a few of them would fall into the margin of the linear-separating hyperplane. As a result, the above constraint can be adjusted according to $y_n(\vec{w}^T\cdot\vec{x}_n+b)\geq 1-\xi_n$, where $\xi_n\geq 0$ for all data points and the total violation is the sum of all $\xi_n$. Using the Kuhn-Tucher theorem, the minimization of  $\frac{1}{2}\vec{w}\cdot\vec{w}$ under the constraints can be achieved by minimizing the following effective Lagrangian,
\begin{eqnarray}
\mathcal{L}(\vec{w},b,\vec{\xi},\vec{\alpha},\vec{\beta})=\frac{1}{2}\vec{w}^T\cdot\vec{w}
+C\sum_{n=1}^N\xi_n\nonumber \\
-\sum_{n=1}^N\alpha_n
[y_n(\vec{w}^T\cdot\vec{x}_n+b)
\nonumber\\
-(1-\xi_n)]
-\sum_{n=1}^N\beta_n\xi_n
\label{eqnarray:Lagrangian}
\end{eqnarray}
where N is the total number of training samples and $\alpha_n, \beta_n\ge 0$ are the Lagrangian multipliers enforcing the constraints. The second term on the r.h.s of Eq.~(\ref{eqnarray:Lagrangian}) is the regularization term that specifies the price that violations of the margin have to pay. Increasing $C$ means less tolerance for violating the margin, thus yields more complex models, whereas decreasing $C$ makes the price of violation smaller, thus avoids overly fitting the noise. The ``hyperparameter'' $C$ should be determined by grid search in a manually specified subset of values. We take the value of $C$ that leads to the best validation result. The validation data samples are generated at values of $\delta J$ in the same range as the testing data set, but for different disorder realizations.

Minimizing $\mathcal{L}$ with respect to $\vec{w}$, $b$, and $\xi_n$ first leads to,
\begin{eqnarray}
\label{eq:4}
\vec{\triangledown}_{\vec{w}}\mathcal{L}=\vec{w}-\sum_{n=1}^N\alpha_ny_n\vec{x}_n=0\\
\frac{\partial\mathcal{L}}{\partial b}=-\sum_{n=1}^N\alpha_ny_n=0\\
\label{eq:6}
\triangledown_{\xi_n}\mathcal{L}=C-\alpha_n-\beta_n=0
\end{eqnarray}
Plugging Eqs~(\ref{eq:4}-\ref{eq:6}) into Eq.~(\ref{eqnarray:Lagrangian}), we can get rid of the variables $\vec{w}$, $\vec{\xi}$, and $b$, and obtain $-\mathcal{L}(\vec{\alpha})$ which is to be minimized with respect to $\vec{\alpha}$:
\begin{equation}
\left.\begin{aligned}
&\frac{1}{2}
\begin{bmatrix}
\alpha_1\\
\alpha_2\\
\vdots\\
\alpha_N
\end{bmatrix}^T
\begin{bmatrix}
y_1y_1\mathcal{K}_{11} \quad &
y_1y_2\mathcal{K}_{12} & \dots &
y_1y_N\mathcal{K}_{1N}\\
y_2y_1\mathcal{K}_{21} \quad &
y_2y_2\mathcal{K}_{22} & \dots &
y_2y_N\mathcal{K}_{2N}\\
\vdots \quad & \vdots & \ddots & \vdots\\
y_Ny_1\mathcal{K}_{N1} \quad &
y_Ny_2\mathcal{K}_{N2} & \dots &
y_Ny_N\mathcal{K}_{NN}\\
\end{bmatrix}
\begin{bmatrix}
\alpha_1\\
\alpha_2\\
\vdots\\
\alpha_N
\end{bmatrix}\\
& -\vec{1}\cdot\vec{\alpha}=-\mathcal{L}(\vec{\alpha})
\end{aligned}\right.
\label{equation:to_minimize}
\end{equation}
under the constraints that  $\sum_{n=1}^N\alpha_ny_n=0$ and $0\leq\alpha_n\leq C, \forall n$, where the $\mathcal{K}_{ij}=\vec{x}_i\cdot\vec{x}_j$ are called the kernel. Note that only a few (out of $N$) of the $\alpha_n$ are nonzero, otherwise there is a high risk for over-fitting. Those nonzero $\alpha_n$ correspond to the data points that are closest to the separating hyperplane. They are the so-called called support vectors because they are what determine the separating hyperplane in the end. After obtaining the $\alpha_n$, $\vec{w}$ can be obtained from Eq.~(\ref{eq:4}) by
\begin{equation}
\vec{w}=\sum_{k=1}^{N_{SV}}\alpha^{SV}_ky^{SV}_k\vec{x}^{SV}_k
\label{eq:compute_w}
\end{equation}
where $\vec{x}^{SV}_k$ is one of the $N_{SV}$ number of the support vectors.

In the above linear algorithm, the kernel $\mathcal{K}_{ij}$ is simply the inner product of two data points $\vec{x}_i$ and $\vec{x}_j$. However, in most of the realistic cases, the data sets are not linearly separable and we have to transform a data point from a vector $\vec{x}$ in its original space $\mathit{X}$ to a vector $\vec{z}$ in a higher dimensional space $\mathit{Z}$. Fig.\ref{fig:schematic}(b) illustrates a simple example of such kind of transformation. If the original $\mathit{X}$ space is 2-dimensional and represented by $(x_1, x_2)$, the simplest transformation to the higher dimensional space   $\mathit{X}\rightarrow\mathit{Z}$ corresponds to be $(x_1,x_2)\rightarrow(x_1^2, \sqrt{2} x_1x_2, x_2^2)$. Consequently, the kernel in $\mathit{Z}$ space is $\mathcal{K}_{ij}=\vec{z}_i\cdot\vec{z}_j=(\vec{x}_i\cdot\vec{x}_j)^2$. In the actual calculations, we only need to know the values of the kernel in order to minimize Eq.~(\ref{equation:to_minimize}) to obtain $\alpha_n$ and thus the decision function.
In fact, a set of input data can be raised to any order by choosing the general form of the polynomial kernel $\mathcal{K}_{ij}=\mathcal{K}(\vec{x}_i,\vec{x}_j)=(c_0+\gamma\vec{x}_i\cdot\vec{x}_j)^d$, or even transformed to infinitely dimensions of space by choosing a radial basis function (RBF) kernel $\mathcal{K}_{ij}=exp(-\gamma|\vec{x}_i-\vec{x}_j|^2)$. The resulting decision function is determined by the value of the kernels according to:
\begin{equation}
\mathit{f}(\vec{x})=\mathrm{sign}\left(\sum_{k=1}^{N_{SV}}\alpha^{SV}_ky^{SV}_k
\mathcal{K}(\vec{x}^{SV}_k,\vec{x})+b\right)
\label{equation:decision_fuc}
\end{equation}
where $\vec{x}^{SV}_k$'s are support vectors.

\section{\label{sec:level3}Phase Classification and Decision Function}

\subsection{Classification result and phase diagram}
In our case, both the training and testing data sets are composed of probability density of the eigenstate wavefunctions of the Hamiltonian in Eq.~(\ref{equation:Hamiltonian}) obtained by exact diagonalization, labeled as MBL ($+1$) or ETH($-1$). We choose $\delta J=0.15\pm 0.05$ and energy densities $\epsilon=59/60$ and $\epsilon=19/60$ which are deep in ETH phase and $\delta J=9.0\pm 1.0$ at the same energy densities which are deep in MBL phase to generate $18000$ wavefunctions, 4500 for each set of $(\delta J, \epsilon)$, and use their probability densities as the training set. We will demonstrate that by training the machine learning models at two different energy densities, the precise values of which are not important, we can obtain a model that works for determining the phase diagram in the whole energy spectrum. More detailed discussion and the possible implications of this remarkable finding will be given at the end of this subsection.

\begin{figure}[h!]
 \includegraphics[trim = 0mm 0mm 0mm 0mm,width=1\columnwidth,clip=true]{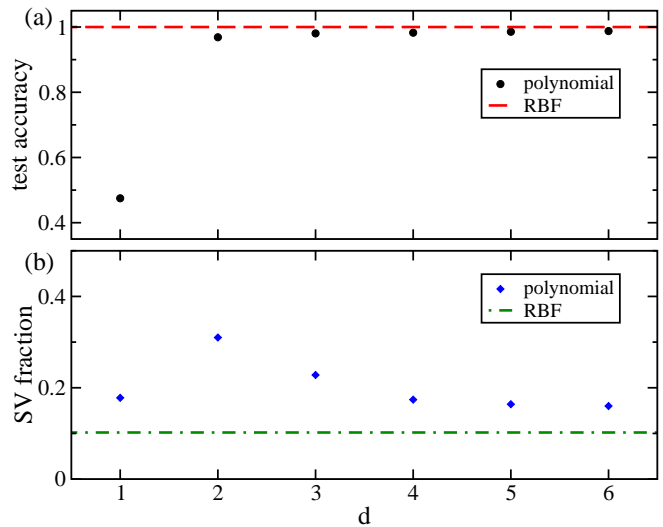}
\caption{(a) The test accuracy as a function of the order ($d$) of the polynomial kernel. The black dots denote the test accuracy. It increases from $47.5\%$ for the linear kernel at $d=1$ and approaches $100\%$ corresponding to that of the RBF kernel (the red dashed line). (b) The fraction of support vectors among all training data versus kernel order $d$ shown in the blue squares. The green dash-dot line corresponds to the fraction of SV in the RBF kernel.  }
\label{fig:model_selection}
\end{figure}
We first train the SVM with different kernels, including the linear kernel, the polynomial kernel with $d=2,3,4,5,6$, and the RBF kernels. Since we only wish to keep the homogeneous terms, we choose $c_0=0$ in $\mathcal{K}_{ij}=(c_0+\gamma\vec{x}_i\cdot\vec{x}_j)^d$ for the polynomial kernels. By grid-search we find that in this case the models are not very sensitive to the regularization. Specifically, when $C$ in Eq.~(2) is swept through $\{10^{-4},10^{-3},10^{-2},10^{-1},1,10,10^2,10^3,10^4\}$, we find that the test accuracy in the validation set always stays above $96\%$ for all polynomial kernels when $C \in [10^{-2}, 10^2]$, which is unaffected by the order $d$ (excluding the special case $d=1$, i.e. the linear kernel). Therefore, we choose $C=1.0$ for our models. For models with polynomial kernels, there exits a threshold of $\gamma$ in the kernel expressions $\mathcal{K}_{ij}=(c_0+\gamma\vec{x}_i\cdot\vec{x}_j)^d$ and $\mathcal{K}_{ij}=exp(-\gamma|\vec{x}_i-\vec{x}_j|^2)$, above which the validation accuracy reaches its maximum. We choose $\gamma=400$, which is large enough to give the optimum validation result for the polynomial models. While for the RBF kernel, we choose $\gamma=1/2^{L+6}$, which is also determined by validation.

Next, we make a model selection of the kernels to adopt based on their performances on the testing set, then use the selected kernel to proceed with the phase classification. The testing set consists of probability density of wavefunctions generated at $\delta J\in [0.05,0.45]$ labeled as ETH and $\delta J\in [9.0,12.0]$ labeled as MBL at $\epsilon=59/60, 43/60, 31/60, 19/60$.  The result for the model selection with $L=12$ is shown in Fig.~\ref{fig:model_selection}(a). We find the test accuracy in the test set is below $50\%$ for the linear kernel, implying that the linear SVM is unable to distinguish between the ETH and MBL phases. The polynomial SVMs, on the otherhand, all have test accuracy above $96\%$, meaning that the polynomial SVMs are all qualified phase classifiers. The test accuracy increases with increasing $d$ until reaching about $100\%$ for the RBF kernel.

\begin{figure}[h!]
 \includegraphics[trim = 0mm 0mm 0mm 0mm,width=1\columnwidth,clip=true]{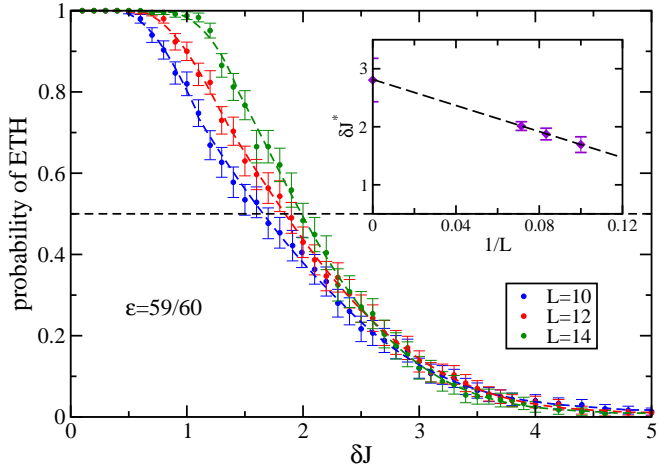}
\caption{
The probability that eigen-wavefunction corresponding to energy density $\epsilon =59/60$ generated at a given $\delta J$ is ETH phase for $\delta J\in [0,5]$. The probability is estimated using the fraction of ETH phase in an ensemble of $300$ disorder realizations at energy density $\epsilon =59/60$ for $L=10$ (blue dots), $L=12$ (red dots) and $L=14$ (red dots) predicted by SVM with RBF kernel. For each size, we take the $\delta J$ corresponding to 50\% probability of being ETH to be the phase boundary and denote it by $\delta J^*$. The inset shows the finite-size extrapolation of $\delta J^*$. The intercept is interpreted as the phase boundary $\delta J_c$ in the thermodynamic limit.}
\label{fig:svm_critical_search}
\end{figure}

In Fig.~\ref{fig:model_selection}(b) we show the fraction of support vectors (SV), namely, the number of nonzero $\alpha_n$ among all training data for $L=12$. The fraction of SV is always smaller than $1/3$. Because the number of SV is directly related to the effective degrees of freedom of the model, this indicates that we are not at the risk of over-fitting. In addition, the fraction of SV decreases with increasing $d$ when $d\ge 2$, until it reaches $10.2\%$ for the RBF kernel. Considering that SV are the data points most difficult to classify, this result again implies that the SVM with the RBF kernel may be the best choice of model for this study. For $L=10$ and $L=14$, the test accuracy versus the order of the polynomial kernel has the same trend as that in $L=12$ case. Thus, we choose the RBF kernel that gives the best test accuracy ($99.81\%$ for $L=10$ and $\sim100\%$ for $L=14$) to search for the phase boundary.

\begin{figure}[h!]
 \includegraphics[trim = 0mm 0mm 0mm 0mm,width=1\columnwidth,clip=true]{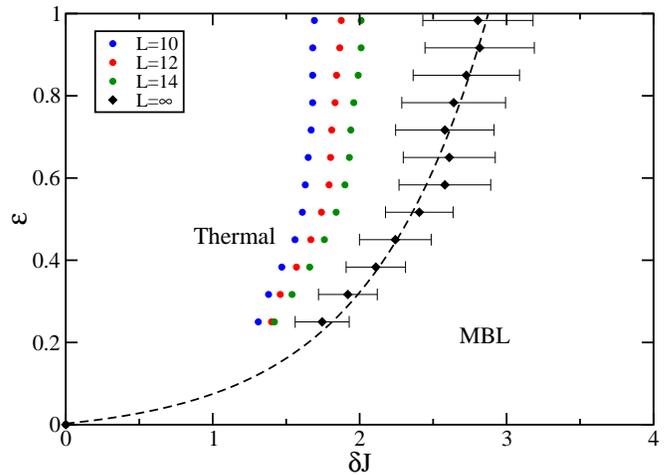}
\caption{Phase diagram of the disordered quantum Ising chain defined in Eq.~(\ref{equation:Hamiltonian}) obtained by SVM with RBF kernel. The training was performed for $0.1\le \delta J\le 0.2$ at two energy densities $\epsilon=59/60$ and $19/60$ labeled as ETH and for $8.0\le\delta J\le 10.0$ at the same two energy densities labeled as MBL. The black diamonds are the critical disorder strengths $\delta J_c$ extracted from the large $L$ extrapolations of the finite size transition points (blue, red and green dots) at the different $\epsilon$. The black dashed line is an exponential fit to the phase boundary.}
\label{fig:svm_phase_diagram}
\end{figure}

Finally, we use the trained SVM to determine the transition point at different energy densities. For better comparison with previous result~\cite{Kjall_2014}, we choose $\epsilon=(11+4i)/60, i=1,2,\cdots,12$. For each of the $\epsilon$, we study a series of $\delta J$ in the range $[0,5]$, and for each $\delta J$ we consider an ensemble of probability density of eigenstate wavefunctions generated with different disorder realizations/configurations. We input all eigenstates in an ensemble and compute the fraction of the ETH outputs. When the ensemble is large enough, this fraction corresponds to the probability that a wavefunction generated at the given $\delta J$ is in ETH phase. The standard deviation of the probability is calculated according to the central limit theorem. The probabilities are plotted with error bar in Fig.~4 as a function of $\delta J$ for different system sizes at a fixed energy density $\epsilon=59/60$. The probability of being in ETH phase behaves like a soft step function. When $\delta J$ is small, namely deep in ETH phase, it approaches 1 because the actual phase should be ETH, whereas for $\delta J$ large, i.e. deep in the MBL phase, it approaches 0. In the transition region between the two limiting phases, the probability of being ETH decreases from $1$ to $0$. We choose the $\delta J$ corresponding to ${\rm ETH \ probability}=0.5$ as the transition point $\delta J^*$ (Fig.~\ref{fig:svm_critical_search}) for a given system size $L$, because it's the disorder strength at which the wavefunctions have half probability to be in ETH phase and half to be in MBL phase thus quantities (like entanglement entropy) that behave differently in these two phases will have the largest standard deviation\cite{Kjall_2014}. As shown in Fig~\ref{fig:svm_critical_search}, with increasing system size, the soft step function becomes steeper, implying that it behaves like a step function in thermal dynamic limit. We regard any disorder strength at which the probability of ETH reaches $0.5$ within error as being in transition region, thus to determine the error of $\delta J^*$. As can be seen from Fig.~4, ${\delta J}^*$ exhibits significant size dependence for $L=14, 12, 10$. In the inset of Fig.~\ref{fig:svm_critical_search}, ${\delta J}^*$ is plotted against $1/L$ and a finite extrapolation within the error bars to the large $L$ limit produces an asymptotic estimate of the $\delta J_c$ separating the ETH and MBL phases at this energy density. Repeating this procedure, we computed $\delta J_c$ at different energy densities $\epsilon=(11+4i)/60, i=1,2,\cdots,12$ shown in the phase diagram Fig.~\ref{fig:svm_phase_diagram}. The phase boundary separating the ETH and the MBL phases is obtained by an exponential fit to the data, which
qualitatively agrees with the result obtained from scaling the variance of the entanglement entropy \cite{Kjall_2014}.

It is important to note that the phase diagram cannot be obtained had the data at only one energy density been used as the training set. Indeed, we started off training the model at a single energy density ($\epsilon=59/60$ or $19/60$) and tested the ability of the model to determine the phase boundaries at different energy densities. Surprisingly, the obtained results were quite poor. The testing accuracy in the best case was below $95\%$. The resulting transition boundary does not vary much with energy and deviates significantly from the one obtained by scaling the variance of the entanglement entropy \citep{Kjall_2014}. This finding is unexpected and remarkable, since it suggests that the information learned by the SVM is controlled by both the energy density and the disorder strength. In order to correctly determine the phase boundary in the two-parameter phase space, the SVM needs to learn to decipher that the information encoded in the wavefunctions come from a two-parameter support in order to avoid being misled by those at different energies. There are at least two possible origins for this novel behavior: (1) this is due to the specifics of the SVM learning algorithm. However, it is worth noting that we find the same property using the neural networks model, which is discussed in detail in the appendix, suggesting that this finding is not specific to a particular machine learning model. It could still arise from the fact that the input to the models, both the training and the processed information, is the probability density of the many-body wavefunctions. (2) An alternative and physically more interesting possibility is that the thermal to MBL transition driven by disorder $\delta J$ and the energy density $\epsilon$ (mobility-edge like) have different critical properties, such that the training along one direction of the phase diagram (at fixed energy density) doesn't enable the model to learn the transition along the other (at fixed disorder). This is reminiscent of the situation where there are two relevant scaling directions at a critical point. Clearly, more works in the future are needed to fully understand this remarkable property.

\subsection{Decision function in SVM}
\begin{figure}[h!]
 \includegraphics[trim = 0mm 0mm 0mm 0mm,width=1\columnwidth,clip=true]{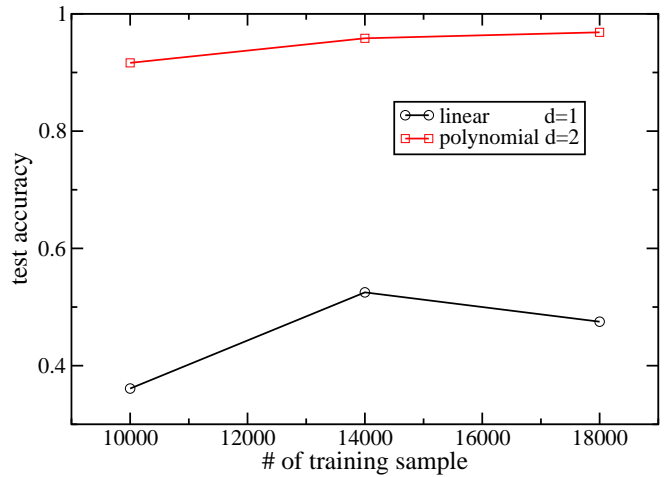}
\caption{The test accuracy obtained from 4500 testing samples of $L=12$ for the SVM machines using the linear kernel (black line) with $\mathcal{K}(\vec{x_i},\vec{x_j})=\vec{x_i}\cdot\vec{x_j}$ and the quadratic polynomial kernel (red line) with $\mathcal{K}(\vec{x_i},\vec{x_j})=(\vec{x_i}\cdot\vec{x_j})^2$. The number of training samples used is indicated by the horizontal axis.
}
\label{fig:svm_acc}
\end{figure}

As can be seen in Fig.~\ref{fig:model_selection}, the linear SVM completely fails to distinguish between the two phases, resulting in  $47.5\%$ test accuracy, in contrast to the worst case of $96.8\%$ for polynomial kernels. We next study the details in the $L=12$ case in order to corroborate our conclusion that the SVM cannot separate the input data labeled by the two different phases in their original space, and that the phase classification requires the transformation of the inputs to higher dimensional spaces. Fig.~\ref{fig:svm_acc} shows that when using linear kernel the test accuracy is around or below $50\%$ in different trials, even with increasing number of training samples. The origin of this can be traced back to the fact that the probability amplitudes of the wavefunctions are normalized so that the sum of elements in an $\vec{x}$, whether they are from the ETH or the MBL regions, is unity. Thus, one can imagine a $2^L-1$ dimensional hyperplane in the feature space where all data samples are distributed because of the constraint. The data points corresponding to MBL phase are more likely to be near the edges of that hyperplane, while the ETH data are more likely to be in the center. It is thus impossible to find a hyperplane of the same dimension to separate them. So we have to turn to at least a quadratic kernel. As shown in Fig.~\ref{fig:svm_acc}, using a quadratic kernel dramatically increases the test accuracy to at least $91.7\%$ with $10000$ training samples, which can be systematically improved further by enlarging the training set. This is what we expect since more training data will reduce model variance, thus improving the test performance.

The unique advantage of the SVM is that one can uncover the exact form of the decision function, although it can be very cumbersome in higher order polynomial kernels and infeasible in the RBF kernel. In the following, we shall limit ourselves to case of the SVM with the quadratic kernel, where the decision function can be written as:

\begin{equation}
\mathit{f}(\vec{z})=\mathrm{sign}\left(\vec{w'}^T\cdot\vec{z}+b\right)=\mathrm{sign}\left(\sum_{i\le j}w'_{ij}z_{ij}+b\right)
\label{equation:decision_func_quad}
\end{equation}
where $i,j=1,2,\cdots,\dim(\mathcal{H})$, and $w'_{ij}$ is each element of $\vec{w'}$ that is coupled to the transformed inputs $\vec{z}$ in the quadratic $\mathit{Z}$ space, according to Eq.~(\ref{eq:compute_w}) it can be calculated as:
\begin{equation}
w'_{ij}=\sum_{k=1}^{N_{SV}}\alpha^{SV}_ky^{SV}_kz^{SV}_{kij}=\sum_{k=1}^{N_{SV}}\alpha^{SV}_ky^{SV}_ku_{ij}x^{SV}_{ki}x^{SV}_{kj}
\label{equation:coefficient}
\end{equation}
given the exact form of the transformation from the original $\mathit{X}$ space to the quadratic $\mathit{Z}$ space: $z_{ij}=u_{ij}x_ix_j$ where $u_{ij}=1$ if $i=j$ and $\sqrt{2}$ if $i<j$. In the same manner, the decision function in Eq.~(\ref{equation:decision_func_quad}) can be written in terms of the original basis as:
\begin{equation}
\mathit{f}(\vec{z})=\mathrm{sign}\left(\sum_{i\le j}w_{ij}x_ix_j+b\right)
\label{equation:decision_func_quad_x}
\end{equation} where $w_{ij}=u_{ij}w'_{ij}$.

\begin{figure}[h!]
 \includegraphics[trim = 0mm 0mm 0mm 0mm,width=1\columnwidth,clip=true]{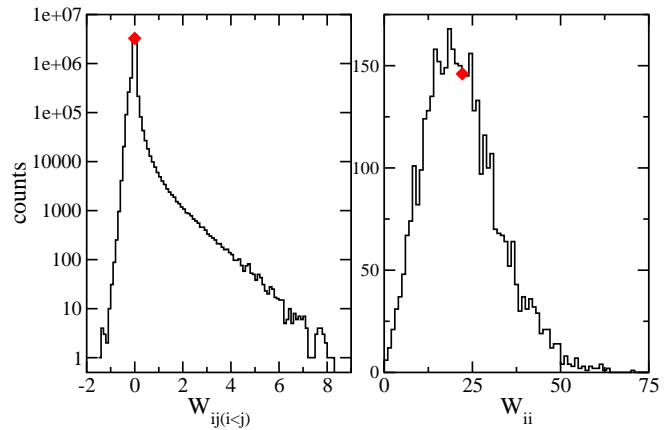}
\caption{Distributions of $W_{ij}$ (left) and $W_{ii}$ (right) for $L=12$. When $i\neq j$, $W_{ij}$ can be positive or negative, but cluster very close to zero with $96.7\%$ of them distributed in the range $[-0.2, 0.2]$ for an average of $-1.8*10^{-3}$ as denoted by the red diamond shown in the left panel. In contrast, the diagonal $W_{ii}$ are much larger. $88.6\%$ of all $W_{ii}$ are larger than $10$ with an average of $22.25$ as denoted by the red diamond in the right panel.}
\label{fig:coef}
\end{figure}

In Fig.~\ref{fig:coef}, we plot the distributions of the off-diagonal and the diagonal values of $w_{ij} (i<j)$ and $w_{ii}$ for $L=12$ where $i,j=1,\cdots,2^{12}$. Clearly, the distributions of $w_{ii}$ and $w_{ij}(i< j)$ are drastically different. We find that $w_{ii}$ coupling to $x_i^2$ are positive for all $i$, with an average of $22.15$, which dominates in the decision function over the contributions from $w_{ij}(i<j)$, which can be either positive or negative but are clustered around much smaller magnitudes with an average of $-1.8*10^{-3}$. As a result, only the diagonal terms of the kind $x_i^2=|\langle\{\sigma_i^z\}|\Psi_n\rangle|^4$ contribute essentially to determining the phase region, whereas the cross term of the form $x_ix_j=|\langle\{\sigma_i^z\}|\Psi_n\rangle|^2 \times|\langle\{\sigma_j^z\}|\Psi_n\rangle|^2$ ($i<j$) do not affect the decision qualitatively. This immediately reminds one of the inverse participation ratio (IPR) that plays a crucial role in the study of the single-particle Anderson localization in disordered media. The generalized definition of the IPR in Fock space of a many-body system is:
\begin{equation}
I_{q}(E_n)=\sum_i|\langle\{\sigma_i^z\}|\Psi_n\rangle|^{2q}
\label{equation:IPR}
\end{equation}
with $q=2$. It can also be seen from Fig.~\ref{fig:coef} that most of the $w_{ii}$ are of the same order, indicating that $\langle\{\sigma_i^z\}|\Psi_n\rangle|^{4}$ for each $i$ contributes almost equally, thus further corroborating that it's a quantity similar to the IPR that acts as the threshold in the decision function of the SVM with the quadratic kernel.

The above analysis and discussion suggest that the decision function of the quadratic SVM is closely related to the many-body IPR $I_{q=2}$. One may wonder if the {\em total} off-diagonal contribution which after averaging over $i$ is $-3.91$, is still negligible compared to the diagonal contribution $x_i^2$ with an average over $i$ being $22.15$. A related question is whether the SVM with higher order polynomial kernels also uses decision functions related to the higher order $I_q$, i.e. if terms like $|\langle\{\sigma_i^z\}|\Psi_n\rangle|^{2q}$ still dominate in the classification for higher $q$. Indeed, Fig.~\ref{fig:model_selection} showed that higher order polynomial kernels lead to better test performance and the test accuracy reaches its maximum for the RBF kernel. It will be instructive to find out the reason for this increase. Is it because the cross terms $x_ix_j$, $i<j$ become more important or more irrelevant, or is it simply because higher order terms are sharper classifiers?

Unfortunately for higher order polynomial kernels, the decision function has poor visualization and becomes even inaccessible in the RBF kernel. So instead of studying the decision functions directly, we preprocess the training data by manually raising each element in the input vector to higher order, removing the cross terms by keeping only terms like $x_i^q=|\langle\{\sigma_i^z\}|\Psi_n\rangle|^{2q}$. Then, we train the linear SVM on the preprocessed data. The test accuracy in the testing set obtained is $99.90\%$ for $q=2$, $99.75\%$ for $q=3$ and $99.69\%$ for $q=4$, suggesting that to correctly distinguish between the MBL and ETH phases, the information from the cross terms are unimportant. Because the test accuracy doesn't change much when varying $q$ in the inputs $|\langle\{\sigma_i^z\}|\Psi_n\rangle|^{2q}$, the  IPR of any order equal to or larger than $2$ can characterize the phase transition. This result also provides a possible explanation for the increase of test accuracy in the higher order polynomial kernels. The contribution from the cross terms to the decision function may be further suppressed in the higher order polynomial and RBF kernels, which causes the test accuracy to approach that obtained without the cross terms.

\begin{figure}[h!]
 \includegraphics[trim = 0mm 0mm 0mm 0mm,width=1\columnwidth,clip=true]{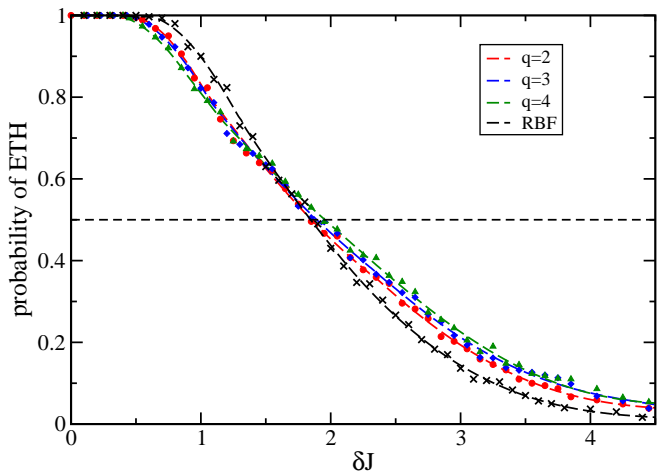}
\caption{Fraction of data points classified as in the ETH phase in an ensemble versus the disorder strength at $\epsilon=59/60$ and $L=12$. The colored symbols and dashed lines denote results obtained by the linear SVM trained on the probability density to the $q$-th power, namely $(x_1^q, x_2^q, \cdots , x_{2^L}^q)$, where $x_i=|\langle\sigma_i^z|\Psi_n\rangle|^2$. Black dashed line is obtained using the RBF kernel trained on the original data set.}
\label{fig:10}
\end{figure}

To gain further insights, we also applied the three linear SVMs trained on the preprocessed data with $q=2,3,4$ to classifying the data in transition region. The results are shown in Fig.~\ref{fig:10} at energy density $\epsilon=59/60$ and $L=12$. The decision boundary obtained in each case corresponds to
$\delta J^*=1.85\pm 0.62$, $1.89\pm 0.65$ and $1.95\pm 0.70$ respectively (shown in colored lines), which agrees well with the result $\delta J^*=1.88\pm 0.47$ for the RBF kernel on original data set (shown in black dashed line). This further supports our conjecture that when the SVMs with polynomial and RBF kernels search for the decision function, they learn to ignore to a large extent the unnecessary cross terms. As before, the decision function of the linear SVM trained on the preprocessed data has contributions from evenly distributed components, $|\langle\{\sigma_i^z\}|\Psi_n\rangle|^{2q}$ with $q=2$, of the same order of magnitude. This is consistent with the decision functions being closely related to the IPR in the spin configuration space.

\subsection{Inverse participation ratio and MBL}

The concept of MBL originates from the inability of many-body eigenstates to thermalize in strongly disordered systems. As such, the entanglement entropy $S_E$ between the subsystems has been the common tool used to separate the ETH phase for weak disorder where $S_E$ obeys the volume-law from a MBL phase at for strong disorder where $S_E$ obeys the area-law and the eigenstates fail to thermalize. There remains under investigation, however, an outstanding issue with important physical implications, i.e. if and how MBL is related to the localization of the eigenstates in the many-body Hilbert-space of the entire system under strong disorder and correlation \cite{Beugeling_2015,Bera_2015,Torres_2016}. 

Our interpretable machine learning results described above have shown that, at least for the disordered quantum spin chain studied, the decision function used by the SVM is related to the generalized many-body IPR in Hilbert space. It is known that relating MBL to the localization in Hilbert space requires a choice of basis and is basis dependent. Because we choose the spin configurations as the basis of the Hilbert space, our SVM approach and its consequent interpretability in terms of IPR is also
specific to this basis. Furthermore, the SVMs can produce the boundary between the ETH and MBL phases, which is in good agreement with the one obtained by scaling the variance of the entanglement entropy\cite{Kjall_2014}, suggesting that the IPR may have the ability to identify the MBL phase transition as a localization phenomenon in the many-body Hilbert space. In single particle picture, the entanglement entropy defined using the site occupation number basis is deterministically related to the IPR and its multifractal spectrum at the Anderson localization transition point \citep{Chen_2012}. Unfortunately, it has not been possible to establish the connection between these two quantities for the many-body eigenstates in disordered interacting systems. Motivated by our machine learning results, in the following, we explore the similarities in the behavior of these two quantities in the disordered quantum spin chain.

\begin{figure}[h!]
 \includegraphics[trim = 0mm 0mm 0mm 0mm,width=1\columnwidth,clip=true]{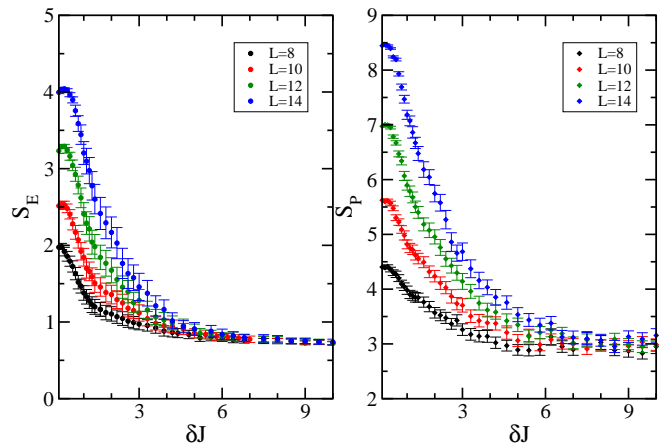}
\caption{The ensemble averaged half-chain entanglement entropy $S_E$ (left panel) and the participation entropy $S_P$ (right panel) plotted versus disorder strength $\delta J$ for different length $L$ of the quantum Ising chain at energy density $\epsilon=59/60$.}
\label{fig:se_sp}
\end{figure}

The entanglement entropy between two partitions separated at the midpoint of the chain is given by, $\mathrm{S_E}=\mathrm{-Tr_L\rho_Lln\rho_L}$, where $\mathrm{\rho_L}$ is the reduced density matrix $\rho_\mathrm{L}=\mathrm{Tr_R}|\Psi\rangle\langle\Psi|$ and $L$ and $R$ denote the left or right half of the chain. In the ETH phase, $S_E$ is an extensive quantity with values proportional to the volume of the subsystem (length $L/2$ of the left half of the chain here) because degrees of freedom in the subsystems are highly entangled. In the MBL phase, however, the entanglement is limited to the boundary between the subsystems such that $S_E$ is proportional to the boundary area. In 1D systems, it is bounded by a constant.

In order to facility a direct comparison to the entanglement entropy $S_E$, we convert the IPR into the participation entropy defined by ${S_P}=\mathrm{-ln}(\sum_i|\langle\{\sigma_i^z\}|\Psi\rangle|^4)$ over the entire system of length $L$. ${S_P}$ is commonly used to study the single-particle Anderson localization \cite{Mirlin_2008,Wegner_1980}. In the single-particle case, when the system is in the delocalized phase, $S_P$ is proportional to the logarithm of the size of configuration space and hence the number of lattice sites in the single-particle picture. In the localized phase, on the other hand, $S_P$ is bounded by a constant. At the mobility edge, i.e. the critical point of the metal-insulator transition, $S_P$ exhibits multifractal behavior. For our interacting Ising chain, the size of the configuration space equals $2^L$. It is thus natural to expect \cite{Luitz_2014,Luitz_2015} $S_P$ to be proportional to the length of the chain $L$ up to certain sub-leading terms in the ETH phase, resembling the volume law behavior of the entanglement entropy $S_E$ in the ETH phase. In the MBL phase, it remains to be explored whether $S_P$ is bounded by a constant, namely, whether there exists a genuine localization in the many-body Hilbert space. We calculate both the entanglement entropy $S_E$ and the participation entropy $S_P$ by exact diagonalization at energy density $\epsilon=59/60$, averaging over ensembles at varying disorder strength $\delta J$. Fig. \ref{fig:se_sp} displays the ensemble averaged $S_E$ (left) and $S_P$ (right) as a function of $\delta J$ for different length of the chain at $L=8,\dots,14$. There are indeed remarkable similarities in their behaviors. At small $\delta J$, both $S_E$ and $S_P$ exhibit clear linear size ($L$) dependence characteristic of the volume-law in the ETH phase. As the $\delta J$ increases, both $S_E$ and $S_P$ decrease, as does their dependence on the system sizes. In the regime of strong disorder with $\delta J$, the entanglement entropy $S_E$ shows essentially no size dependence, characteristic of a MBL phase with the area-law in 1D. The participation entropy $S_P$ also displays a much reduced size-dependence, which disappears for the largest sample sizes $L=10$ and $12$ at large disorder $\delta J$. While a definitive conclusions would require numerical studies of even larger system sizes which are beyond our current size limit, these results together with those from the interpretable machine learning studies
bring sufficient new insights and raise the possibility of studying theoretically as well as experimentally\cite{Roushan_2017} other physical quantities more directly connected to the localization of the many-body eigenstates in the Hilbert space.

\section{\label{sec:level5}Summary and Conclusions}

We presented in this paper an interpretable machine learning classification of the thermal and MBL phases in a disordered quantum Ising spin chain. Specifically, the SVMs were built with different types of kernels of the probability density of the exact eigenstate wavefunctions. We find that training the machines with data at a minimal of two different energy densities and two disorder strengths corresponding to the limiting cases deep in the thermal and MBL phases, the SVMs are able to classify the phases in the {\em entire} transition regime and determine the boundary separating the two phases at {\em all} energy densities. The phase boundary determined by machine learning is in good qualitative agreement with that obtained by scaling the variance of the entanglement entropy\cite{Kjall_2014}. These results show that the decision function of the SVM is a general two-parameter quantity, i.e. the energy density and disorder strength, capable of classifying the whole many-body eigenstate spectrum of the Hamiltonian. In addition to providing insights into the critical behavior of the MBL transition, these findings also demonstrate the the efficiency of machine learning classification in that it can operate with much less labeled data which are expensive in computation. Thus, when appropriately applied, the SVMs can be more powerful tool for classifying physical data compared to conventional methods, especially in complex physical situations. In the appendix, we also trained the 3-layer neural networks (NN) machine on the same training sets, and used it in the same way as the SVM to classify the MBL and thermal phases. The phase diagram obtained by neural network machine agrees to that determined by the SVM within the error bars, demonstrating that different machine learning models lead to the consistent classification results in the disordered quantum Ising spin chain.

A unique advantage of the SVM is its interpretability, which indeed allowed us to interpret how the SVM separates the input data belonging to the different phases. Remarkably, we find that the decision function constructed by the SVM is closely related to the generalized IPR in the many-body Hilbert space. The fact that the interpretable machine learning suggests that IPR may have the ability to identity the MBL transition is a physically significant results in that it relates the failure to thermalize to the Anderson type of localization in the many-body Hilbert space. The consistency between the SVM phase diagram and the one obtained from the variance of the entanglement entropy\cite{Kjall_2014} further supports this intriguing possibility. Introducing the participation entropy to describe the many-body IPR, we further explored this connection by directly comparing the entanglement entropy and the participation entropy and found remarkable similarity in their behaviors. Further studies of the interconnection between these two quantities in larger system sizes are however necessary to reach more definitive conclusions.

\section{acknowledgements}

We thank Yi Zhang, Sen Zhou and Xiuzhe Luo for helpful discussion. The work is supported in part by the U.S. Department of Energy, Basic Energy Sciences Grant No. DE-FG02-99ER45747 (W.Z. and Z.W.). Z.W. thanks the hospitality of the Institute of Physics, the Chinese Academy of Sciences. L.W. acknowledge support by the the National Natural Science Foundation of China under Grant No. 11774398.

\appendix*

\section{Phase classification by artificial neural networks}

The artificial neural networks (NN) are computing systems widely used in data classification, pattern recognition an so on. Recently its application to condensed matter physics has been explored heavily,  with significant outcomes.\cite{Carrasquilla_2017,Nieuwenburg_2017,Carleo_2017,Schindler_2017} Here we train a one-hidden-layer NN on the same training data used in the main text, namely, the probability density of wave functions generated at small disorder strength $\delta J=0.15\pm 0.05$ at $\epsilon=19/60, 59/60$, labeled as ETH, denoted by a $2$ dimensional vector $(1,0)$ and probability density of wave functions generated at large $\delta J=9.0\pm 1.0$ at the same energy densities, labeled as MBL, denoted by $(0,1)$.

\begin{figure}[ht]
\centering
\includegraphics[scale=0.4]{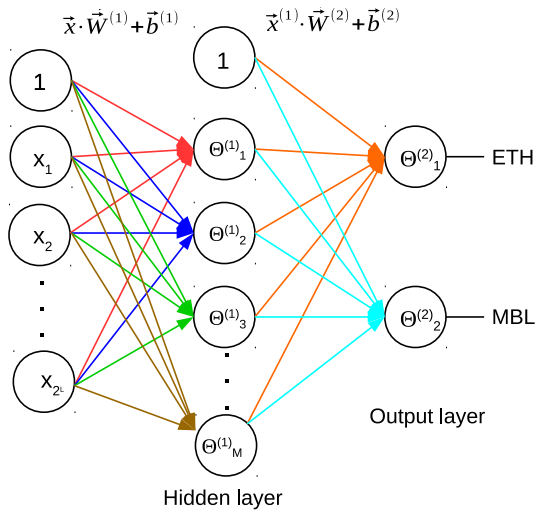}
\caption{Schematic explanation how NN maps an input data $\vec{x}_i$ to its label $y_i$, the NN acts on all input data points $i=1,2,\cdots, N$ thus plays a role as its target function.}
\label{fig:1}
\end{figure}

\begin{figure}[ht]
\centering
\includegraphics[width=\linewidth]{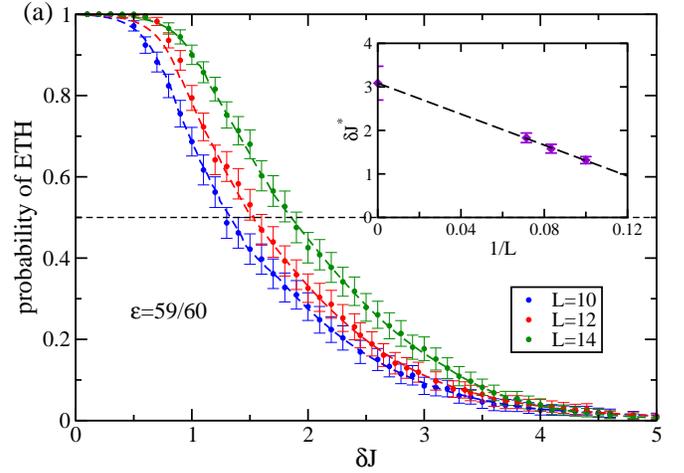}
\caption{The probability that eigen-wavefunction corresponding to energy density $\epsilon =59/60$ generated at a given $\delta J$ is ETH phase for $\delta J\in [0,5]$. The probability is estimated by the fraction of ETH phase in an ensemble of $300$ disorder realizations at energy density $\epsilon =59/60$ for $L=10$ (blue dots), $L=12$ (red dots) and $L=14$ (red dots) predicted by NN. For each size, we take the $\delta J$ corresponding to 50\% probability of being ETH to be the phase boundary and denote it by $\delta J^*$. The inset shows the finite-size extrapolation of $\delta J^*$. The intercept is interpreted as the phase boundary $\delta J_c$ in the thermodynamic limit.
}
\label{fig:2}
\end{figure}

In the hidden layer, the inputs $\vec{x}$ are multiplied by a $2^L\times M$ dimensional matrix $\vec{\vec{W}}^{(1)}$, where $M$ is the number of nodes in the hidden layer, $M$ ranges from $80$ to $200$ depending on the dimension of inputs, in another word, the size $L$ of the spin chain. After the above linear combination, the results are added to some biases $\vec{b}^{(1)}$ and then fed to a nonlinear activation function $\Theta^{(1)}$. The work done by the first layer can be summarized as:
\begin{equation}
\vec{x}^{(1)}=\Theta^{(1)} (\vec{x}\cdot\vec{\vec{W}}^{(1)}+\vec{b}^{(1)})
\label{equation:1}
\end{equation}
where $\Theta^{(1)}$ takes the form of ReLU\cite{bishop}, and $\vec{x}^{(1)}$ are the outputs of the hidden layer.

Similarly, the next layer, called the output layer, maps $\vec{x}^{(1)}$ to the final outputs $f(\vec{x})$ by
\begin{equation}
f(\vec{x})=\Theta^{(2)} (\vec{x}^{(1)}\cdot\vec{\vec{W}}^{(2)}+\vec{b}^{(2)})
\label{equation:2}
\end{equation}
where $\vec{W}^{(2)}$ is a $M\times 2$ dimensional vector performing linear combination of $\vec{x}^{(1)}$, $\vec{b}^{(2)}$ are the biases, and $\Theta^{(2)}$ takes the form of softmax\cite{bishop} function. Thus the NN maps the $2^L$ dimensional inputs to $2$ dimensional outputs, Fig.~\ref{fig:1} illustrates how NN works. The two elements of a $2$ dimensional output represent the probability that the input being classified as ETH and being classified as MBL respectively. The final prediction of class should be the class whose probability is larger in $f(\vec{x})$.

We use cross entropy as the cost function that acts as a metric to describe the closeness between the outputs $f(\vec{x})$ and the actual labels $\vec{y}$.
\begin{equation}
Cost=-\sum_{i=1}^N \vec{y}_i\cdot\log f(\vec{x}_i)
\label{equatio:3}
\end{equation}
where $\vec{x}_i$ denotes input of each training sample and $\vec{y}_i$ denotes the corresponding label, $N$ is the total number of training samples. All parameters of NN, including $\vec{\vec{W}}^{(1)}, \vec{\vec{W}}^{(2)}, \vec{b}^{(1)}, \vec{b}^{(2)}$, are determined by minimizing the cost function.

\begin{figure}[htbp]
\centering
\includegraphics[width=\linewidth]{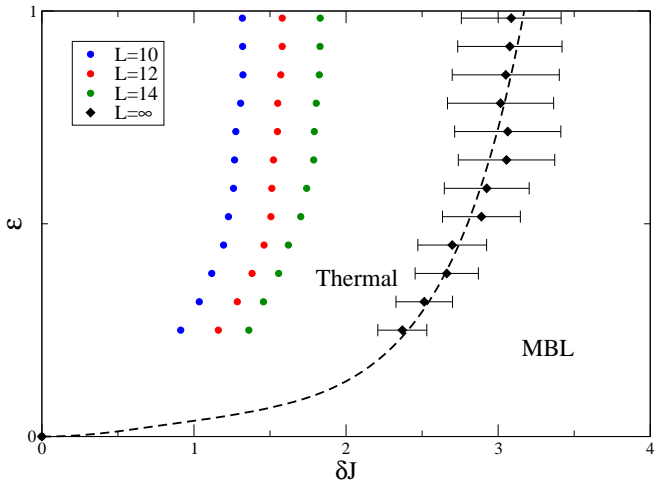}
\caption{Phase diagram of the disordered quantum Ising chain defined in main text. With $\epsilon=2(E-E_{min})/(E_{max}-E{min}$) being the energy density relative to the total bandwidth. The black diamonds are $\delta J_c$ at different $\epsilon$s, which are the finite size extrapolations from the finite size transition point (blue, red and green dots). }
\label{fig:3}
\end{figure}

We use the same testing set as that used by SVM described in the main text. The testing accuracy is $99.8\%$ with $L=14$, accuracy $99.5\%$ with $L=12$, and accuracy $98.8\%$ with $L=10$. We then follow the same procedure described in the main text to determine critical points for energy densities $\epsilon=(11+4i)/60, i=1,2,\cdots, 12$  (Fig.~\ref{fig:2}), and then the phase boundary separating MBL and thermal phases by exponential fitting(Fig.~\ref{fig:3}). The result obtained by using NN agrees with that of SVM within error, it also agrees with that of scaling the variance of entanglement entropy\cite{Kjall_2014} within error.

\bibliography{ml-mbl}

\end{document}